\documentclass[preprint]{aastex}

\usepackage{amsmath,amssymb}
\usepackage{natbib}
\usepackage{multirow}
\usepackage{url}
\usepackage{lscape}
\defcitealias{YangX12}{Paper I}

\hfuzz=\maxdimen
\shorttitle{Magnetic Nonpotentiality as Predictor of Flares}
\shortauthors{YANG ET AL.}

\begin{document}
%\begin{CJK}{GBK}{song}

\title{Magnetic Nonpotentiality in Photospheric Active Regions \\ as a Predictor of Solar Flares}

\author{Xiao~YANG\altaffilmark{1,2}, GangHua~LIN\altaffilmark{1}, HongQi~ZHANG\altaffilmark{1}, and XinJie~MAO\altaffilmark{1,3}}
\email{yangx@nao.cas.cn}
\altaffiltext{1}{Key Laboratory of Solar Activity, National Astronomical Observatories, Chinese Academy of Sciences, 20A Datun Road, Beijing 100012, P. R. China}
\altaffiltext{2}{Graduate University of Chinese Academy of Sciences, No.19A Yuquan Road, Beijing 100049, P. R. China}
\altaffiltext{3}{Department of Astronomy, Beijing Normal University, No.19 Xinjiekouwai Street, Beijing 100875, P. R. China}

\begin{abstract}
  Based on several magnetic nonpotentiality parameters obtained from the vector photospheric active region magnetograms obtained with the Solar Magnetic Field Telescope at the Huairou Solar Observing Station over two solar cycles, a machine learning model has been constructed to predict the occurrence of flares in the corresponding active region within a certain time window. The Support Vector Classifier, a widely used general classifier, is applied to build and test the prediction models. Several classical verification measures are adopted to assess the quality of the predictions. We investigate different flare levels within various time windows, and thus it is possible to estimate the rough classes and erupting times of flares for particular active regions. Several combinations of predictors have been tested in the experiments. The True Skill Statistics are higher than 0.36 in 97\% of cases and the Heidke Skill Scores range from 0.23 to 0.48. The predictors derived from longitudinal magnetic fields do perform well, however they are less sensitive in predicting large flares. Employing the nonpotentiality predictors from vector fields improves the performance of predicting large flares of magnitude $\geq$M5.0 and $\geq$X1.0.
\end{abstract}


\section{INTRODUCTION}

Solar flares are sudden processes that release tremendous energy in a short period of time in the solar atmosphere. They lead to transient heating of local regions and the dramatic enhancement of electromagnetic radiation and high-energy particle ejection. Some large eruptions toward the Earth have an impact on normal human activities. It is worthwhile to make short-term predictions of solar flares to reduce losses. For a long period of time, solar physicists have been trying to understand the physics of flares, in order to make predictions by simulating the evolutions of magnetic fields in the solar atmosphere and by obtaining information from the solar interior. At present, however, it seems relatively feasible to make predictions based on the statistical relationships between solar eruptions and the evolution of other solar phenomena. Some authors predict flares based on morphological parameters or remote information from different sources (e.g., \citealt{Gallagher02,Qahwaji07,LiR07,Colak09,Bloomfield12}). Such predictors require manual intervention before entering the prediction process, and therefore are not suitable for automatic operations. There are some other flare-prediction studies adopting the measures deduced from longitudinal magnetic fields (e.g., \citealt{Georgoulis07,YuDR09,SongH09,Mason10,YuanY10,Ahmed11}).

The accumulation of magnetic nonpotential energy is of importance for solar eruptions. \citet{Mason10} mentioned the importance of the vector-field data to obtain the most promising flare-predictive magnetic parameters. Leka and Barnes (\citeyear{Leka07}) contributed a great amount to the exploration of the differences of magnetic-field properties between flare-imminent and flare-quiet active regions, however, the number or the time spans of their samples were quite restricted. Lacking long-term consistent observations of vector magnetic fields, the magnetic nonpotentiality was rarely used in solar flare predictions. The vector magnetograms obtained at the Huairou Solar Observing Station over more than 20 yr make the experiments possible. \citeauthor{YangX12} (\citeyear{YangX12}, hereafter \citetalias{YangX12}) have calculated the statistical relations between magnetic nonpotentiality and solar flares. By means of the prediction experiments described in this Letter, we can predict the occurrence of flares in particular active regions based on their magnetic properties alone, and also can estimate the starting time and eruption magnitude of the flares. In addition, several classical verification measures of dichotomous predictions are discussed to call for more serious concerns on the verification issue \citep{Doswell90}. The Heidke Skill Scores (HSS) and the True Skill Statistics (TSS; see Section~\ref{S-SS}) of our 100 group experiments are in the ranges 0.23--0.48 and 0.32--0.82, respectively. Our results show that the nonpotentiality predictors improve the performance of predicting more powerful flares.

\section{DATA AND METHOD}
\label{S-DataMethod}

  \subsection{Data and Preprocessing}

We use the observational data of photospheric active region vector magnetograms obtained by the Solar Magnetic Field Telescope (SMFT; \citealt{AiGX86}) at the Huairou Solar Observing Station, National Astronomical Observatories of China. SMFT is a 35 cm aperture vector magnetograph with a tunable birefringent filter. The working spectral line for the vector magnetograms is Fe\,{\small I} $\lambda$5324.19, which is a strong and broad line with an equivalent width of about 0.334 \textrm{\AA} and a Land\'{e} factor of 1.5 (\citealt{AiGX82}). The data employed are selected from all the vector magnetograms during the period from 1988 to 2008 subject to the following criteria: (1) the active regions are located within 30\,$^{\circ}$ from the solar disk center, and (2) only one magnetogram is used for each active region in one observation day. The final data set, which is also used in \citetalias{YangX12}, consists of 2173 photospheric vector magnetograms involving 1106 active regions. The detailed descriptions of the data and their distributions during the two solar cycles are in \citetalias{YangX12}, as well as the calibration for the vector magnetograms and the determination of the 180\,$^{\circ}$ ambiguity of the transverse field. The records of soft X-ray flares are available from NOAA's National Geophysical Data Center.\footnote{\url{ftp://ftp.ngdc.noaa.gov/STP/space-weather/solar-data/solar-features/solar-flares/x-rays/goes/}}

  \subsection{Magnetic Nonpotentiality Parameters as Predictors}
            \label{S-predictor}

The magnetic nonpotentiality parameters as predictors, the inputs for the prediction model, are the mean planar magnetic shear angle $\overline{\Delta\phi}$, mean shear angle of the vector magnetic field $\overline{\Delta\psi}$, mean absolute vertical current density $\overline{|J_{z}|}$, mean absolute current helicity density $\overline{|h_{\rm c}|}$, absolute averaged twist force-free factor $|\alpha_{\rm av}|$, mean free magnetic energy density $\overline{\rho_{\rm free}}$, effective distance of the longitudinal magnetic field $d_{\rm E}$, longitudinal-field weighted effective distance $d_{\rm Em}$ \citepalias{YangX12}, mean horizontal gradient of the longitudinal field $\overline{\nabla_{\rm h}B_{z}}$, maximum horizontal gradient $(\nabla_{\rm h}B_{z})_{\rm m}$, length of strong-gradient ($>$0.05 G km$^{-1}$) inversion lines $L_{\rm gnl}$, and mean density of longitudinal magnetic energy dissipation $\overline{\varepsilon(B_{z})}$ (\citealt{CuiYM06,JingJ06}). All of the above measures are macroscopic and averaged quantities, which indicate the magnetic nonpotentiality or magnetic complexity of a whole active region. In the calculations, each magnetogram is represented as ($\mathbf{x}_{i}$,$y_{i}$), where $\mathbf{x}_{i}\in \mathbf{R}^{n}$ is the predictor array and $y_{i}\in \{1,-1\}$ is the class label of the magnetogram ($y_{i}=1$ for flaring instances and $y_{i}=-1$ for non-flaring ones, according to the labeling scheme stated in Section~\ref{S-ExpDes}). We have tried five combinations of predictors:
\begin{description} \small
\item
V06 ($\overline{\Delta\psi}$, $\overline{|J_{z}|}$, $\overline{|h_{\rm c}|}$, $|\alpha_{\rm av}|$, $\overline{\rho_{\rm free}}$, $d_{\rm Em}$),
\item
V08 ($\overline{\Delta\phi}$, $\overline{\Delta\psi}$, $\overline{|J_{z}|}$, $\overline{|h_{\rm c}|}$, $|\alpha_{\rm av}|$, $\overline{\rho_{\rm free}}$, $d_{\rm E}$, $d_{\rm Em}$),
\item
L05 ($d_{\rm Em}$, $\overline{\nabla_{\rm h}B_{z}}$, $(\nabla_{\rm h}B_{z})_{\rm m}$, $L_{\rm gnl}$, $\overline{\varepsilon(B_{z})}$),
\item
A10 ($\overline{\Delta\psi}$, $\overline{|J_{z}|}$, $\overline{|h_{\rm c}|}$, $|\alpha_{\rm av}|$, $\overline{\rho_{\rm free}}$, $d_{\rm Em}$, $\overline{\nabla_{\rm h}B_{z}}$, $(\nabla_{\rm h}B_{z})_{\rm m}$, $L_{\rm gnl}$, $\overline{\varepsilon(B_{z})}$),
\item
A12 ($\overline{\Delta\phi}$, $\overline{\Delta\psi}$, $\overline{|J_{z}|}$, $\overline{|h_{\rm c}|}$, $|\alpha_{\rm av}|$, $\overline{\rho_{\rm free}}$, $d_{\rm E}$, $d_{\rm Em}$, $\overline{\nabla_{\rm h}B_{z}}$, $(\nabla_{\rm h}B_{z})_{\rm m}$, $L_{\rm gnl}$, $\overline{\varepsilon(B_{z})}$).
\end{description}

  \subsection{Prediction Method: Support Vector Classification}

Predicting whether or not an active region will flare within a certain time interval can be transformed into a classification problem. The support vector machine (SVM) first introduced by Vapnik (\citealt{Boser92,Cortes95,Vapnik95}) is now a widely applied statistical learning theory used to solve classification and regression problems. In recent years, SVM has been applied to the field of astronomy (e.g., \citealt{ZhangYX03,Wozniak04,Wadadekar05,GaoD08,Beaumont11,PengNB12}) including solar physics (e.g., \citealt{QuM03,Qahwaji07,LiR07,AlOmari10,Labrosse10,Alipour12}). A machine learning system for classification is able to learn and construct a model (from the existing training data with definite category labels) which can classify the training data and predict upcoming ones whose categories are unknown. The maximum margin principle and the kernel function are the two core concepts of the SVM. By solving an optimization problem, the SVM classifier is obtained as an optimal separating hyperplane $\mathbf{w}\cdot\mathbf{x}+b=0$ that separates the two-class data with the maximum distance. When in a linearly non-separable case, a kernel function is employed, then the training vectors $\mathbf{x}_{i}$ are mapped into a higher-dimensional feature space in which the data can be linearly separated.

The primal optimization problem can be written as
\begin{eqnarray*}
\min_{\mathbf{w}\in\mathcal{H},b\in\mathbf{R},\boldsymbol{\xi}\in\mathbf{R}^{l}} && \frac{1}{2} \|\mathbf{w}\|^{2}+C\sum_{i=1}^{l} \xi_{i}\, , \nonumber \\
\textrm{subject to} && y_{i}(\mathbf{w}\cdot\phi(\mathbf{x}_{i})+b)\geqslant 1-\xi_{i}\, ,\,i=1,\cdots,l\, , \nonumber \\
 && \xi_{i}\geqslant0\, ,\,i=1,\cdots,l\, . \nonumber
\end{eqnarray*}
$C>0$ is the penalty parameter for the sum of slack variables $\xi_{i}$. $\frac{1}{2}\|\mathbf{w}\|^{2}$ corresponds to the distance maximization of the two classes. Taking the reciprocal, the square, and the factor 1/2 are for mathematical convenience. $\phi(\mathbf{x}_{i})$ denote the training vectors in the higher-dimensional space after employing the kernel function. The kernel function is denoted by $K(\mathbf{x}_{i},\mathbf{x}_{j})$, and the corresponding dual optimization problem, which is easier to solve, is
\begin{eqnarray*}
\min_{\boldsymbol{\alpha}} && \frac{1}{2}\sum_{i=1}^{l}\sum_{j=1}^{l}y_{i}y_{j}\alpha_{i}\alpha_{j}K(\mathbf{x}_{i},\mathbf{x}_{j})-\sum_{j=1}^{l}\alpha_{j}\, , \nonumber \\
\textrm{subject to} && \sum_{i=1}^{l}y_{i}\alpha_{i}=0\, , \nonumber \\
 && 0\leqslant\alpha_{i}\leqslant C,\, i=1,\cdots,l, \nonumber
\end{eqnarray*}
where $\alpha_{i}$ are the Lagrange multipliers. Then the coefficients ${\alpha_{i}}^{*}$ for the optimal hyperplane are solved from the dual problem. The training vectors $\mathbf{x}_{i}$ with $\alpha_{i}^{*}\neq0$ are the support vectors that contribute to the final discriminant function
\begin{eqnarray*}
f(\mathbf{x})=\mathrm{sgn}(\mathbf{w}^{*}\cdot\phi(\mathbf{x})+b^{*})=\mathrm{sgn}\left(\sum_{i=1}^{l} \alpha_{i}^{*}y_{i}K(\mathbf{x},\mathbf{x}_{i})+b^{*}\right),
\end{eqnarray*}
where $\mathbf{w}^{*}$ and $b^{*}$ are the corresponding solutions of the primal problem ($b^{*}=y_{j}-\sum_{i=1}^{l}y_{i}\alpha_{i}^{*}K(\mathbf{x}_{i},\mathbf{x}_{j})$ taking any $0<\alpha_{j}^{*}<C$). The plus and minus signs of $f(\mathbf{x})$ indicate the two different classes.

There are a few commonly used kernels like the polynomial kernel, Gaussian radial basis kernel, sigmoid kernel, etc. After trying several kernels in the calculations, we accept the Gaussian radial basis kernel, the mathematical expression of which is $K(\mathbf{x}_{i},\mathbf{x}_{j})=\exp(-\|\mathbf{x}_{i}-\mathbf{x}_{j}\|^{2}/\sigma^{2})$, where $\sigma$ is a kernel parameter. The SVM software LIBSVM (\citealt{ChangCC11}) is used in our experiments.

Note that this classification or prediction model is based on statistical relations with no obvious physical meanings; nevertheless, the physical parameters closely related to solar flares must make positive effects to the performance of the model. This is exactly why we adopt magnetic nonpotentiality and complexity parameters as predictors.

\section{EXPERIMENTS AND RESULTS}
\label{S-Result}
  \subsection{Experiment Design} \label{S-ExpDes}
According to whether the active regions produce flares exceeding a specified class within a certain time window, every magnetogram is labeled as positive (flaring) or negative (non-flaring). The ``time window'' in this Letter begins at the observing time of each magnetogram. We set the flaring magnitude thresholds to C1.0, C5.0, M1.0, M5.0, and X1.0, and the time windows 6, 12, 24, and 48 hr. The positive--negative sample ratios are different for different combinations of flaring thresholds and time windows (see Table~\ref{T-sample}). In each labeled set, we divide the whole set into training and testing subsets, then train the training subset to obtain the classifier and test the rest to evaluate the performance of the classifier.

\begin{table}[ht] \footnotesize
\begin{center}
\caption{Flaring (f) and Non-flaring (n-f) Sample Distributions}
\label{T-sample}
\begin{tabular}{ccccccc}
\noalign{\smallskip}\hline
\hline\noalign{\smallskip}
 & & \multicolumn{5}{c}{Flaring Threshold} \\
\cline{3-7}\noalign{\smallskip}
Time Window & Category & C1.0 & C5.0 & M1.0 & M5.0 & X1.0 \\
\hline\noalign{\smallskip}
\multirow{2}{*}{48 hr} & f & 918 & 427 & 252 & 71 & 42 \\
 & n-f & 1255 & 1746 & 1921 & 2102 & 2131 \\
\hline\noalign{\smallskip}
\multirow{2}{*}{24 hr} & f & 697 & 291 & 167 & 39 & 25 \\
 & n-f & 1476 & 1882 & 2006 & 2134 & 2148 \\
\hline\noalign{\smallskip}
\multirow{2}{*}{12 hr} & f & 475 & 181 & 95 & 22 & 17 \\
 & n-f & 1698 & 1992 & 2078 & 2151 & 2156 \\
\hline\noalign{\smallskip}
\multirow{2}{*}{6 hr} & f & 309 & 101 & 58 & 12 & 9 \\
 & n-f & 1864 & 2072 & 2115 & 2161 & 2164 \\
\hline
\end{tabular}
\end{center}
\end{table}

$k$-fold cross-validation is used for avoiding overfitting. The full set is randomly divided into $k$ subsets with approximately equal size, $(k-1)$ of which are for training and the remaining is for testing. Training and testing are repeated $k$ times. Each subset is tested exactly once. We take $k=10$ for most sets, and $k=5$ for the sets whose flaring samples are less than 50. The positive--negative sample ratios of both training and testing subsets are maintained consistent with that of the original set.

  \subsection{Performance Assessment for Predictions} \label{S-SS}

The counts of successes and failures obtained from previous dichotomous prediction constitute a 2$\times$2 contingency table (confusion matrix in machine learning), as shown in Table \ref{T-conmat}. The verification measures assessing the prediction performance are derived from the statistics in the table. For simplicity, we use the notations ($x$,$y$,$z$,$w$) to name the four elements of the contingency table. $x$ is the number of the positive events predicted positive (True Positive or Hit), $y$ the number of the positive events predicted negative (False Negative or Miss), $z$ the number of the negative events predicted positive (False Positive or False Alarm), and $w$ the number of the negative events predicted negative (True Negative or Correct Rejection). $x$ and $w$ make a positive impact on the prediction assessment while $y$ and $z$ do the opposite.

\begin{table}[ht] \footnotesize
\begin{center}
\caption{Definition of the 2$\times$2 Contingency Table (Confusion Matrix)}
\label{T-conmat}
\begin{tabular}{ccccl}
\noalign{\smallskip}\hline
\hline\noalign{\smallskip}
 & \multicolumn{3}{c}{Predicted} & \\
\cline{2-4}\noalign{\smallskip}
Observed & Yes & \ \ & No & Total \\
\hline\noalign{\smallskip}
 Yes & $x$ & \ \ & $y$ & $N_{1}=x+y$ \\
 No & $z$ & \ \ & $w$ & $N_{0}=z+w$ \\
\hline\noalign{\smallskip}
 & & & & $N=x+y+z+w$ \\
\hline

\end{tabular}
\end{center}
\end{table}

From Table~\ref{T-conmat}, we can directly obtain eight ratios of the elements with their associated marginal sums: POD\footnote{Probability of Detection.}$=x/(x+y)$, FOH\footnote{Frequency of Hits, also named true positive rate.}$=x/(x+z)$, FAR\footnote{False Alarm Ratio.}$=z/(x+z)$, POFD\footnote{Probability of False Detection.}$=z/(z+w)$, FOM\footnote{Frequency of Misses.}$=y/(x+y)$, DFR\footnote{Detection Failure Ratio.}$=y/(y+w)$, PON\footnote{Probability of a Null event.}$=w/(z+w)$, and FOCN\footnote{Frequency of Correct Null forecasts, also named true negative rate.}$=w/(y+w)$ (cf. \citealt{Doswell90}). POD, FOH, PON, and FOCN, in which the numerator is $x$ or $w$, are hoped to be higher, and the other four are expected to be lower. Other verification measures are also available such as $F_{1}$-measure, HSS, TSS, Critical Success Index (CSI), Gilbert Skill Score (GSS), and Clayton Skill Score (CSS), a summary of which is shown in Table \ref{T-SS}. The perfect prediction, which is difficult to achieve in practice, corresponds to these verification measures reaching their upper bounds of 1. Though it has been more than a century since the ``Finley affair'' (see \citealt{Murphy96}) inducing hot discussions, the study on this seemingly simple 2$\times$2 problem remains ongoing (\citealt{Stephenson00}). In this work, we only consider the classical verification measures which are more intuitional to utilize in practical operations.

The percentage of correct predictions $(x+w)/N$ (referred as ACC hereafter) is the simplest but often misleading measure to assess the prediction, especially when one side, event or non-event, is overwhelming. ACC, CSI, and $F_{1}$ do not exclude the correct numbers based on the stochastic prediction. The so-called skill scores indicate the relative accuracy of a prediction to some standard reference predictions. The generic form of skill score is
\begin{eqnarray*}
{\rm SS}=\frac{S-S_{\rm ref}}{S_{\rm perfect}-S_{\rm ref}}\times 100\%,
\end{eqnarray*}
where $S$ is a particular measure of accuracy, $S_{\rm ref}$ a reference, and $S_{\rm perfect}$ the perfect prediction. A no-skill prediction scores 0, a positive score shows a better prediction than the reference, and the perfect prediction scores 1. HSS is a skill score from ACC comparing with the random prediction. GSS is the skill-modified CSI, subtracting the expected correct predictions due to chance from $x$. $F_{1}$ is the harmonic mean of POD and FOH, and HSS happens to be the harmonic mean of skill-modified POD and skill-modified FOH (${\rm POD}_{\rm s}$ and ${\rm RS}_{\rm s}$ in \citealt{Schaefer90}). The skill-modified ones are always lower than the original ones.

These verification measures are related to each other through the connections of $x$, $y$, $z$, and $w$. A common property of HSS, GSS, TSS, and CSS is that they all have the factor $(x\cdot w-y\cdot z)$ in their numerators. This factor becomes zero in the random prediction, and thus these four skill scores all have the value 0, indicating no skill. In the constant prediction (all positive predictions, $y=w=0$; or all negative, $x=z=0$), this factor is also zero; CSS is meaningless in this case. The values of CSI, $F_{1}$, and ACC in random situations depend on the ratio of events to non-events. Another common property of the above four skill scores is that they are all fair to both events and non-events. Considering non-events as focus, swapping $x$ with $w$ and $y$ with $z$ simultaneously, they remain unchanged; this is not the case with CSI or $F_{1}$.

Keeping the numerators of the above four skill scores exactly the same, the differences of their denominators are:
\begin{eqnarray*}
D_{\rm HSS}-D_{\rm TSS} &=& \frac{1}{2}(y-z)(y-z+x-w), \\
D_{\rm TSS}-D_{\rm CSS} &=& (z-y)(x-w), \\
D_{\rm GSS}-D_{\rm HSS} &=& \frac{1}{2}(y+z)(x+y+z+w), \\
D_{\rm GSS}-D_{\rm TSS} &=& y(x+y)+z(z+w).
\end{eqnarray*}
GSS is usually less than TSS and HSS, except when $y=z=0$ (i.e., in the perfect prediction). There is no definite magnitude relation between TSS and CSS, or between HSS and TSS. HSS is less than TSS if $w$ is overwhelmingly dominant ($w\gg x$ and usually $z>y$ in optimizing TSS). There is little difference between HSS and TSS if $w$ is not dominant. Detailed introductions to the contingency table and forecast verification can be found in \citet{Wilks06}.

  \subsection{Experiment Results and More Comments on Verification}

It is nearly impossible to optimize all the verification measures simultaneously (\citealt{Manzato05}; see also the results of \citealt{Bloomfield12}). Accordingly, we compute the geometric mean of several verification measures (POD, FOH, TSS, HSS, GSS, CSI, $F_{1}$, and $\sqrt[3]{{\rm POD}\cdot {\rm FOH} \cdot {\rm FOCN}}$) which we are more concerned about. A grid search process is carried out to obtain a relatively better pair of $(C,\sigma^{2})$ for the final SVM classifier. A12's results\footnote{The complete results are available in \url{http://sun.bao.ac.cn/~yangx/files/yx_NonpPred_result.pdf}.} are shown in Table~\ref{T-result}, in which each value with its error is the arithmetical mean of the specific verification measure in $k$ times testing. The percentage of non-events ($N_{0}/N$) is given at the end of each row for reference. $F_{1}$ is always higher than CSI, except when $x=0$ or $y=z=0$. In rare event situations, HSS is close to $F_{1}$, so HSS is likely higher than CSI. In our results, there are only two cases with HSS lower than CSI (C1.0, 48 hr; C1.0, 24 hr). These are the top two cases whose positive samples are in a larger proportion compared with other cases and $w$ is not extremely dominant. The predictors derived from longitudinal magnetic fields (L05) perform somewhat better than those mainly involving transverse components (V06, V08) in predicting flares of $\geq$C1.0 and $\geq$C5.0. However, the superiority diminishes in predicting more powerful flares. For instance, in the case of $\geq$M5.0 or $\geq$X1.0 flares, the performance of longitudinal predictors becomes worse than that of other predictor combinations. It seems that the predictors from longitudinal fields are less sensitive in predicting large flares. Overall, there is an improvement in the prediction employing various measures derived from vector magnetic fields (A10, A12).

HSS and TSS are often discussed and applied in forecast verification (e.g., \citealt{Woodcock76,Doswell90,Manzato05}). ${\rm HSS}={\rm TSS}$ when $y=z$; ${\rm HSS}\equiv {\rm TSS}$ when $N_{1}=N_{0}$. \citet{Bloomfield12} proposed using TSS instead of HSS as a standard to reliably compare flare forecasts. However, no single scalar measure can cover all the information of the prediction results. Even the unbiased TSS, which is independent of the event frequency, fails to effectively deal with rare event predictions (\citealt{Doswell90}). TSS approaches POD in rare event situations, so both $w$ and $z$ contribute little to the results. Experientially, $z$ rises if $x$'s proportion is increased. The bias ($(x+z)/(x+y)\neq 1$) may be unintentionally introduced in optimizing a verification measure \citep{Manzato05}. Pursuing higher POD or TSS will cause higher FAR and lower FOH. Fewer misses cost more false alarms, but ``crying wolf'' may be undesirable. Moreover, the same TSS does not mean the same prediction performance. For instance, Table \ref{T-example} lists some examples from \citet{Woodcock76}. The prediction P1 has POD = 75\% and PON = 50\%, and P2 has POD = 50\% and PON = 75\%. TSS remains the same in the two cases and two predictions, but the results are indeed different. Therefore, only one measure might mislead the prediction verification, and \textit{multiple} verification measures are probably acceptable. This point of view is as well mentioned in \citet{Schaefer90}, \citet{Doswell90}, \citet{Marzban98}, etc. We believe that, since each data set may have its own intrinsic properties, it is inappropriate to compare different predictions on different trial samples.

\section{CONCLUSIONS AND DISCUSSIONS}
\label{S-ConDis}

Based on the long-term reliable observations of the photospheric vector magnetic fields by SMFT, we adopt some nonpotentiality measures which are not available from observations of only line-of-sight magnetic fields to study the prediction of solar flares. Real-time processing and no manual intervention are two advantages of our prediction system. The data for the input of the prediction model are obtained by local observations, and the key measures as predictors are available without manual operations.

From our experiments, the combinations of magnetic measures derived from longitudinal fields perform well in the flare prediction, however, they may be less sensitive than the measures from vector fields in predicting large flares. The information of transverse fields makes a limited contribution to the prediction of low magnitude flares, but it does improve the prediction for large flares such as $\geq$M5.0 and $\geq$X1.0 ones. Thus, it is reasonable to include transverse field components in flare predictions.

To avoid misleading the optimization work or misusing the results from a single verification measure, prediction results should be assessed carefully. It is helpful to consider \textit{multiple} verification measures. A step like $k$-fold cross-validation is necessary for improving the generalization capability of the prediction models. The intrinsic properties of various data sets may make a specific tool perform rather differently, and hence, it is then significant to make comparisons in the same data environment.

Some researchers have begun to use vector magnetograms from the Helioseismic Magnetic Imager (HMI) on board the \textit{Solar Dynamics Observatory} to predict solar flares. Yet, the prediction methods founded on statistical information are restricted by the finite time span of HMI data at present. Results of statistical predictions depend on both the historical data set and prediction method employed. There is still a long way to go for the prediction of solar activities employing the exquisite HMI data.

\acknowledgments

The authors are very grateful to the anonymous referees for encouraging comments and beneficial suggestions that improved the manuscript. We are indebted to the HSOS staff and the \textit{GOES} team for the data they produced. X. Y. acknowledges helpful discussions with Dr. Shangbin Yang. This work is supported by the National Natural Science Foundation of China (60940030, 10921303, 41174153, 10903015, 11003025, 11103037, 11103038, 11203036, and 11221063), the Young Researcher Grant of National Astronomical Observatories of Chinese Academy of Sciences (CAS), the Knowledge Innovation Program of CAS (KJCX2-EW-T07), the National Basic Research Program of MOST (2011CB811401), and the Key Laboratory of Solar Activity of CAS.

%\bibliographystyle{apj}
%\bibliography{yx_NonpPred_bib}

\begin{table} \renewcommand{\arraystretch}{2.0} \small
  \begin{center}
  \caption{Verification Measures (VM)}
  \label{T-SS}
  \begin{tabular}{clllr}
  \noalign{\smallskip}\hline
  \hline
  VM & Derivation & Formulation & $w$-Dominated & Range \\
  \hline\noalign{\smallskip}
  GSS\tablenotemark{a} & ${\rm GSS}=\frac{x-E_{1}\tablenotemark{g}}{(x-E_{1})+y+z}$ & ${\rm GSS}=\frac{x\cdot w-y\cdot z}{(y+z)(x+y+z+w)+x\cdot w-y\cdot z}$ & $\rightarrow$ CSI & $[-1/3,1]$\\
  HSS\tablenotemark{b} & ${\rm HSS}=\frac{(x-E_{1})-(w-E_{0}\tablenotemark{h})}{(x-E_{1})+y+z+(w-E_{0})}$ & ${\rm HSS}=\frac{2(x\cdot w-y\cdot z)}{(x+y)(y+w)+(x+z)(z+w)}$ & $\rightarrow F_{1}$ & $[-1,1]$\\
  TSS\tablenotemark{c} & ${\rm TSS}={\rm POD}-{\rm POFD}$ & ${\rm TSS}=\frac{x\cdot w-y\cdot z}{(x+y)(z+w)}$ & $\rightarrow$ POD & $[-1,1]$\\
  CSS\tablenotemark{d} & ${\rm CSS}={\rm FOH}-{\rm DFR}$ & ${\rm CSS}=\frac{x\cdot w-y\cdot z}{(x+z)(y+w)}$ & $\rightarrow$ FOH & $[-1,1]$\\
  CSI\tablenotemark{e} & ${\rm CSI}=\frac{x+w-w}{x+y+z+w-w}$ & ${\rm CSI}=\frac{x}{x+y+z}$ & CSI & $[0,1]$ \\
  $F_{1}$\tablenotemark{f} & $F_{1}=2({\rm POD}^{-1}+{\rm FOH}^{-1})^{-1}$ & $F_{1}=\frac{2x}{2x+y+z}$ & $F_{1}$ & $[0,1]$\\
  \hline
  \end{tabular}
  \tablenotetext{a}{Gilbert Skill Score (\citealt{Gilbert1884}; see \citealt{Schaefer90}).}
  \tablenotetext{b}{Heidke Skill Score (\citealt{Doolittle1888,Heidke26}; see \citealt{Woodcock76,Doswell90}).}
  \tablenotetext{c}{True Skill Statistic, also called Peirce Skill Score or Hanssen-Kuipers' discriminant (\citealt{Peirce1884,Hanssen65}; see \citealt{Woodcock76,Doswell90}).}
  \tablenotetext{d}{Critical Success Index, also called threat score (\citealt{Gilbert1884,Donaldson75}; see \citealt{Schaefer90}).}
  \tablenotetext{e}{Clayton Skill Score (\citealt{Clayton34}; see \citealt{Wandishin02}).}
  \tablenotetext{f}{$F_{\beta}$ measure, $\beta=1$ (\citealt{Van79,Chinchor92}).}
  \tablenotetext{g}{$E_{1}=(x+z)(x+y)/N$, the expected number of correct event predictions due to chance.}
  \tablenotetext{h}{$E_{0}=(y+w)(z+w)/N$, the expected number of correct non-event predictions due to chance.}
  \end{center}
\end{table}

\begin{table} \small
\begin{center}
\caption{An Example of Two Predictions on Two Cases}
\label{T-example}
\begin{tabular}{cccccccc}
\noalign{\smallskip}\hline
\hline\noalign{\smallskip}
\multicolumn{7}{l}{Case 1: $N_{1}=x+y=140,\ N_{0}=z+w=60$} \\
\hline\noalign{\smallskip}
$(x,\ y,\ z,\ w)$ \ \ \ & TSS & HSS & GSS & CSS & \ \ \ CSI & $F_{1}$ \\
\noalign{\smallskip}\hline\noalign{\smallskip}
P1: (105, 35, 30, 30) \ \ \ & 0.250 & 0.244 & 0.139 & 0.239 & \ \ \ 0.617 & 0.764 \\
\noalign{\smallskip} P2: (70, 70, 15, 45) \ \ \ & 0.250 & 0.198 & 0.110 & 0.215 & \ \ \ 0.452 & 0.622 \\
\noalign{\smallskip}\hline
\hline\noalign{\smallskip}
\multicolumn{7}{l}{Case 2: $N_{1}=x+y=60,\ N_{0}=z+w=140$} \\
\hline\noalign{\smallskip}
$(x,\ y,\ z,\ w)$ \ \ \ & TSS & HSS & GSS & CSS & \ \ \ CSI & $F_{1}$ \\
\noalign{\smallskip}\hline\noalign{\smallskip}
P1: (45, 15, 70, 70) \ \ \ & 0.250 & 0.198 & 0.110 & 0.215 & \ \ \ 0.346 & 0.514 \\
\noalign{\smallskip} P2: (30, 30, 35, 105) \ \ \ & 0.250 & 0.244 & 0.139 & 0.239 & \ \ \ 0.316 & 0.480 \\
\noalign{\smallskip}\hline
\end{tabular}
\end{center}
\end{table}

\clearpage
\thispagestyle{empty}
\begin{landscape}
\begin{table} \footnotesize \addtolength{\tabcolsep}{-3pt}
\begin{center}
\caption{Verification Results from Testing SVM Classifier}
\label{T-result}
%\begin{rotate}{270}
\begin{tabular}{cclllllllllll}
\noalign{\smallskip}\hline
\hline\noalign{\smallskip}
\multicolumn{2}{c}{Prediction} & \multicolumn{8}{c}{Verification Measure} & \multicolumn{1}{c}{}\\
\cline{1-12}\noalign{\smallskip}
Flare Level & Time Window & POD & FOH & POCN & CSI & $F_{1}$ & TSS & CSS & HSS & GSS & ACC & $N_{0}/N$ \\
\hline\noalign{\smallskip}
\multirow{4}{*}{$\geqslant$C1.0} & 48 hr & 0.707$\pm$0.011 & 0.690$\pm$0.013 & 0.782$\pm$0.008 & 0.538$\pm$0.013 & 0.698$\pm$0.011 & 0.474$\pm$0.020 & 0.472$\pm$0.020 & 0.473$\pm$0.020 & 0.312$\pm$0.018 & 0.742$\pm$0.010 & 0.578 \\
 & 24 hr & 0.677$\pm$0.019 & 0.617$\pm$0.013 & 0.840$\pm$0.008 & 0.478$\pm$0.016 & 0.645$\pm$0.014 & 0.478$\pm$0.022 & 0.458$\pm$0.020 & 0.466$\pm$0.021 & 0.306$\pm$0.018 & 0.761$\pm$0.010 & 0.679 \\
 & 12 hr & 0.653$\pm$0.028 & 0.508$\pm$0.010 & 0.895$\pm$0.007 & 0.399$\pm$0.013 & 0.569$\pm$0.014 & 0.475$\pm$0.024 & 0.404$\pm$0.014 & 0.430$\pm$0.016 & 0.275$\pm$0.013 & 0.786$\pm$0.006 & 0.781 \\
 & 6 hr & 0.560$\pm$0.026 & 0.423$\pm$0.022 & 0.923$\pm$0.004 & 0.317$\pm$0.017 & 0.479$\pm$0.020 & 0.430$\pm$0.026 & 0.346$\pm$0.025 & 0.378$\pm$0.024 & 0.235$\pm$0.018 & 0.826$\pm$0.009 & 0.858 \\
\hline\noalign{\smallskip}
\multirow{4}{*}{$\geqslant$C5.0} & 48 hr & 0.627$\pm$0.026 & 0.549$\pm$0.019 & 0.906$\pm$0.005 & 0.415$\pm$0.020 & 0.584$\pm$0.020 & 0.500$\pm$0.027 & 0.455$\pm$0.024 & 0.474$\pm$0.025 & 0.313$\pm$0.021 & 0.825$\pm$0.008 & 0.803 \\
 & 24 hr & 0.626$\pm$0.028 & 0.457$\pm$0.022 & 0.939$\pm$0.004 & 0.357$\pm$0.018 & 0.524$\pm$0.018 & 0.507$\pm$0.026 & 0.396$\pm$0.024 & 0.437$\pm$0.022 & 0.282$\pm$0.019 & 0.847$\pm$0.008 & 0.866 \\
 & 12 hr & 0.485$\pm$0.044 & 0.400$\pm$0.028 & 0.952$\pm$0.004 & 0.283$\pm$0.028 & 0.435$\pm$0.033 & 0.419$\pm$0.044 & 0.352$\pm$0.031 & 0.379$\pm$0.035 & 0.239$\pm$0.029 & 0.896$\pm$0.006 & 0.917 \\
 & 6 hr & 0.595$\pm$0.064 & 0.250$\pm$0.029 & 0.979$\pm$0.003 & 0.219$\pm$0.029 & 0.351$\pm$0.038 & 0.508$\pm$0.065 & 0.229$\pm$0.032 & 0.306$\pm$0.041 & 0.187$\pm$0.029 & 0.898$\pm$0.006 & 0.954 \\
\hline\noalign{\smallskip}
\multirow{4}{*}{$\geqslant$M1.0} & 48 hr & 0.642$\pm$0.028 & 0.433$\pm$0.021 & 0.950$\pm$0.004 & 0.350$\pm$0.021 & 0.516$\pm$0.023 & 0.531$\pm$0.030 & 0.383$\pm$0.024 & 0.438$\pm$0.026 & 0.284$\pm$0.021 & 0.860$\pm$0.007 & 0.884 \\
 & 24 hr & 0.550$\pm$0.039 & 0.419$\pm$0.023 & 0.962$\pm$0.003 & 0.314$\pm$0.023 & 0.474$\pm$0.028 & 0.486$\pm$0.039 & 0.381$\pm$0.026 & 0.424$\pm$0.030 & 0.273$\pm$0.023 & 0.907$\pm$0.004 & 0.923 \\
 & 12 hr & 0.554$\pm$0.056 & 0.344$\pm$0.021 & 0.979$\pm$0.002 & 0.266$\pm$0.025 & 0.415$\pm$0.030 & 0.505$\pm$0.052 & 0.323$\pm$0.022& 0.382$\pm$0.031 & 0.240$\pm$0.024 & 0.934$\pm$0.004 & 0.956 \\
 & 6 hr & 0.523$\pm$0.067 & 0.225$\pm$0.028 & 0.986$\pm$0.002 & 0.191$\pm$0.028 & 0.312$\pm$0.039 & 0.474$\pm$0.067 & 0.211$\pm$0.030 & 0.286$\pm$0.040 & 0.173$\pm$0.028 & 0.939$\pm$0.004 & 0.973 \\
\hline\noalign{\smallskip}
\multirow{4}{*}{$\geqslant$M5.0} & 48 hr & 0.634$\pm$0.048 & 0.319$\pm$0.020 & 0.987$\pm$0.002 & 0.268$\pm$0.020 & 0.419$\pm$0.025 & 0.587$\pm$0.047 & 0.306$\pm$0.021 & 0.393$\pm$0.026 & 0.247$\pm$0.020 & 0.942$\pm$0.004 & 0.967 \\
 & 24 hr & 0.329$\pm$0.114 & 0.434$\pm$0.094 & 0.988$\pm$0.002 & 0.231$\pm$0.075 & 0.353$\pm$0.093 & 0.321$\pm$0.114 & 0.422$\pm$0.095 & 0.343$\pm$0.094 & 0.224$\pm$0.075 & 0.980$\pm$0.003 & 0.982 \\
 & 12 hr & 0.460$\pm$0.104 & 0.275$\pm$0.067 & 0.994$\pm$0.001 & 0.213$\pm$0.056 & 0.338$\pm$0.075 & 0.447$\pm$0.105 & 0.269$\pm$0.067 & 0.329$\pm$0.076 & 0.207$\pm$0.056 & 0.982$\pm$0.002 & 0.990 \\
 & 6 hr & 0.667$\pm$0.139 & 0.220$\pm$0.038 & 0.998$\pm$0.001 & 0.202$\pm$0.042 & 0.327$\pm$0.058 & 0.654$\pm$0.139 & 0.218$\pm$0.039 & 0.322$\pm$0.059 & 0.198$\pm$0.042 & 0.985$\pm$0.002 & 0.994 \\
\hline\noalign{\smallskip}
\multirow{4}{*}{$\geqslant$X1.0} & 48 hr & 0.522$\pm$0.070 & 0.487$\pm$0.141 & 0.991$\pm$0.001 & 0.326$\pm$0.072 & 0.474$\pm$0.082 & 0.507$\pm$0.070 & 0.477$\pm$0.141 & 0.462$\pm$0.084 & 0.316$\pm$0.073 & 0.976$\pm$0.005 & 0.981 \\
 & 24 hr & 0.480$\pm$0.102 & 0.375$\pm$0.094 & 0.994$\pm$0.001 & 0.262$\pm$0.061 & 0.401$\pm$0.078 & 0.469$\pm$0.102 & 0.369$\pm$0.095 & 0.392$\pm$0.079 & 0.256$\pm$0.061 & 0.983$\pm$0.003 & 0.982 \\
 & 12 hr & 0.533$\pm$0.062 & 0.278$\pm$0.023 & 0.996$\pm$0.001 & 0.214$\pm$0.010 & 0.353$\pm$0.013 & 0.522$\pm$0.061 & 0.274$\pm$0.023 & 0.346$\pm$0.013 & 0.210$\pm$0.010 & 0.985$\pm$0.002 & 0.990 \\
 & 6 hr & 0.700$\pm$0.200 & 0.169$\pm$0.055 & 0.999$\pm$0.001 & 0.167$\pm$0.056 & 0.270$\pm$0.084 & 0.688$\pm$0.199 & 0.167$\pm$0.056 & 0.265$\pm$0.084 & 0.164$\pm$0.056 & 0.987$\pm$0.003 & 0.996 \\
\hline
\end{tabular}
\end{center}
\end{table}
\end{landscape}
\clearpage

%\end{CJK}
\end{document}